\documentstyle[prl,twocolumn,aps,floats]{revtex} 
\begin{document}

\bibliographystyle{prsty}

\noindent{\bf Spin Tunneling in Mn-12 Cannot be Assisted by Phonons at Millikelvin
Temperatures [Comment on PRL ${\bf 83}$, 416 (1999) by Bellessa {\it et al}]}\\
\\
Recently, Bellessa {\it et al} \cite{Bellessa} in a Letter entitled ``Phonon-Assisted
Tunneling in High-Spin Molecules: Experimental Evidence'', reported observation of a peak in the imaginary part of the ac-susceptibility ${\chi}''$ of Mn-12 molecular nanomagnet, that decreases strongly below 0.1\,K down to 0.02\,K. They explained this effect by a ``phonon-induced tunneling process''. The purpose of this Comment is to show
that the explanation of Bellessa {\it et al} is logically inconsistent and physically impossible. All formulas of Ref.\  \cite{Bellessa} supporting that explanation are incorrect.

The Letter of Bellessa {\it et al} reports observation of the resonant absorption of
the energy of the ac magnetic field of frequency ${\omega}_{0}$ at the value of the transverse magnetic field, $H_{\perp}{\approx}$6\,T, that satisfies ${\hbar}{\omega}_{0}={\Delta}(H_{\perp}$), where ${\Delta}$ is the tunneling splitting
of the ground state. The corresponding peak in the dependence of ${\chi}''$ on 
$H_{\perp}$ first grows as the temperature is lowered down to 0.1\,K but then decreases
down to 0.02\,K. The authors of Ref.\  \cite{Bellessa} give the following explanation to
this effect. Citing p.\,122 of Abragam and Bleaney \cite{AB}, they write ${\chi}''$
as  
\begin{displaymath}
{\rm Eq}.\,(3)\!: \;\;{\chi}'' = CN{\Delta}^{2}f({\omega})T_{2}\tanh({\hbar}{\omega}/2k_{B}T) 
\;. 
\end{displaymath}
According to Bellessa {\it et al}, ``$C$ is a constant, $N$ is the number of spins, and ${\Delta}$ is the magnetic dipole matrix element between the two states of the
fundamental dublet. The shape function $f({\omega})$ is a Lorentzian function of the applied frequency ${\omega}$, the resonance frequency ${\omega}_{0}$ (which depends on the applied magnetic field), and the relaxation time $T_{2}$ describing the linewidth.'' Right after that statement the authors of Ref.\  \cite{Bellessa} go on saying ``We explain our effect by assuming that the tunneling rate ${\Delta}$ in Eq.\ (3) is induced by a two-phonon process: the magnetic moment makes a transition from $+|\psi>$ to $-|\psi>$ (or vice versa) and absorbs (or emits) a quantum ${\hbar}{\omega}$ from the ac magnetic field only if a phonon of angular frequency ${\omega}$ is absorbed and then reemitted after the transition. This process is quite similar to the Raman process, except that the frequencies of the two phonons are the same''. To account for this effect, Bellessa {\it et al} simply insert into Eq.\ (3) $n(n+1)$, where $n=n({\omega}/T)$ is the phonon occupation number. This gives them the desired decrease of ${\chi}''$ at low temperature. The absence of this effect above 0.1\,K is blamed on the unknown temperature dependence of $T_{2}$. 

To begin with, how can ${\Delta}$ be ``the magnetic dipole matrix element'' and ``the tunneling splitting'' at the same time? According to the trivial calculation of
Abragam and Bleaney \cite{AB}, Eq.\ (3) must contain the dipole matrix element
$|{\mu}|^{2}=(g{\mu}_{B}S)^{2}$ ($S=10$ being the spin of the Mn-12 molecule) instead of ${\Delta}^{2}$. This equation describes solely the resonant absorption by Mn-12 molecules of photons of frequency ${\omega}$ generated by the ac-field. Why should that process, in the millikelvin range, be accompanied by the absorption and re-emission of phonons? If it was true, Eq.\ (3), besides the factor $n(n+1)$, must have been multiplied by the fourth power of ratio of the matrix element of spin-phonon coupling
to the Debye temperature, $|V_{s-ph}/(k_{B}{\Theta}_{D})|^{4}$, which is a very small number. In addition, it would have been multiplied by the phase volume of the two phonons, which is nearly zero if ``the frequency is the same''. ``Same frequency'',
in fact, has nothing to do with the Raman process of emission and absorption of two real phonons satisfying ${\hbar}({\omega}_{1}-{\omega}_{2})={\Delta}$, with 
${\hbar}{\omega}_{1}{\sim}{\hbar}{\omega}_{2}{\sim}k_{B}T$. This latter process should be absolutely negligible in the millikelvin range as compared to the direct photon absorption given by Eq.\ (3). The suggestion of Ref.\ \cite{Bellessa} is, therefore, total absurd. 

It is not the purpose of this Comment to speculate why the resonance value of ${\chi}''$ goes down below 0.1\,K. Nevertheless, assuming that the experiment is correct, we will suggest two possibilities. The first is the onset of magnetic ordering in a crystal of Mn-12 clusters due to magnetic dipole interaction between the clusters \cite{MCA}. The second is the ordering of nuclear spins of Mn atoms inside the cluster \cite{GCS}. Both types of ordering induce an effective field acting on the cluster. The bias induced by that field drives the cluster off resonance, preventing it from tunneling. Some support to this suggestion
comes from the fact that 0.1\,K is the right order of magnitude for both dipole ordering
temperature and nuclear ordering temperature \cite{MCA,GCS}.  

This work has been supported by the NSF Grant No. DMR-9978882.\\

%%%%%%%%%%%%%%%%%%%%%%%%%%%%%%%%%%%%%%%%%%%%%%%%%%%%%%%%%%% \vspace{0.5cm}
\noindent
E. M. Chudnovsky and D. A. Garanin\\
Physics Department, CUNY Lehman College\\ Bedford Park Boulevard West, Bronx, NY 10468-1589\\ 
chudnov@lehman.cuny.edu\\
%%%%%%%%%%%%%%%%%%%%%%%%%%%%%%%%%%%%%%%%%%%%%%%%%%%%%%%%%%% %\begin{thebibliography}{99}


\begin{references}
\bibitem{Bellessa} G. Bellessa, N. Vernier, B. Barbara, and D. Gatteschi, Phys. Rev. Lett. {\bf 83}, 416 (1999).
\bibitem{AB} A. Abragam and B. Bleaney, Electron Paramagnetic Resonance of
Transition Ions (Clarendon Press, Oxford, 1970).
\bibitem{MCA} X. Martinez-Hidalgo, E. M. Chudnovsky, and A. Aharony,
cond-mat/0004215 (unpublished).
\bibitem{GCS} D. A. Garanin, E. M. Chudnovsky, and R. Schilling, Phys. Rev.
{\bf B61}, 12204 (2000).
%\end{thebibliography}
\end{references}
\end{document}